\documentclass[useAMS,usenatbib]{mn2e}
\usepackage{graphicx}

\title[A relativistic ejection in XTE~J1550$-$564 in 1998]{Revisiting the relativistic ejection event in XTE~J1550$-$564 during the 1998 outburst}
\author[Hannikainen et al.]{D.C.~Hannikainen$^1$$\thanks{E-mail:
diana@kurp.hut.fi}$
R.W.~Hunstead$^2$,
K. Wu$^3$,
V. McIntyre$^4$, 
J.E.J. Lovell$^5$, \and 
D. Campbell-Wilson$^2$, 
M.L. McCollough$^6$, 
J. Reynolds$^4$ and A.K. Tzioumis$^4$ \\
$^1$ Mets\"ahovi Radio Observatory, TKK, Mets\"ahovintie 114, FI-02540 Kylm\"al\"a, Finland\\
$^2$ Sydney Institute for Astronomy, School of Physics, University of Sydney, NSW 2006, Australia\\
$^3$ Mullard Space Science Laboratory, University College London, Holmbury St Mary, 
   Dorking, Surrey RH5 6NT, UK\\
$^4$ ATNF, CSIRO, PO Box 76, Epping, NSW 1710, Australia\\
$^5$ School of Mathematics \& Physics, Private Bag 21, University of Tasmania, Hobart TAS 7001, Australia \\
$^6$ Harvard-Smithsonian Center for Astrophysics, 60 Garden Street, Cambridge, MA 02138, USA } 

\date{Received: }
  
\begin{document}

\def\Mdot{\hbox{$\dot M$}}
\def\Msun{\hbox{M$_\odot$}}
\def\Rwd{\hbox{$R_{_{\rm WD}}$\,}}
\def\GS{\lower.5ex\hbox{$\buildrel>\over\sim$}}
\def\LS{\lower.5ex\hbox{$\buildrel<\over\sim$}}
\def\kms{km ${\rm s}^{-1}$}    
\def\xte1550{XTE~J1550$-$564}

\maketitle

\begin{abstract}
We revisit the discovery outburst of the X-ray transient \xte1550 during 
   which relativistic jets were observed in 1998 September, and review the 
   radio images obtained with the Australian Long Baseline Array,
   and lightcurves obtained with the Molonglo Observatory Synthesis
   Telescope and the Australia Telescope Compact Array. 
Based on H{\sc i} spectra, we constrain the source distance to
   between 3.3 and 4.9~kpc. 
The radio images, taken some two days apart, show the evolution of an ejection event. 
The apparent separation velocity of the two outermost ejecta is 
  at least $1.3c$ and may be as large as $1.9c$;
  when relativistic effects are taken into account, the inferred true
  velocity is $\geq0.8c$.
  The flux densities appear to peak simultaneously during the outburst, with a
  rather flat (although still optically thin) spectral index of $-0.2$.
\end{abstract}


\section{Introduction}\label{intro}

The soft X-ray transient \xte1550 was discovered by the All-Sky Monitor (ASM) 
  on board the {\it Rossi X-ray Timing Explorer} ({\it RXTE}) 
  on MJD~51063 (1998 Sep 7; MJD=JD$-$240000.5) 
  with an intensity of $\sim$0.07~Crab in the 2--12~keV range (Smith 1998)
  and by the Burst and Transient Source Experiment (BATSE) 
  on the {\it Compton Gamma-Ray Observatory} ({\it CGRO}).
The flux in the 20--100~keV range 
  showed an impulsive rise, 
  increasing from $<0.1 \times 10^{-8} \ \rm erg \ {\rm  s}^{-1} \ {\rm cm}^{-2}$ 
  to $2.42\times 10^{-8} \ \rm erg \ {\rm  s}^{-1} \ {\rm cm}^{-2}$ 
  between MJD~51063 and 51066 (Wilson et al. 1998). 
The {\it RXTE}/ASM intensity increased steadily over the next few days, 
  reaching $\sim$1.7~Crab on MJD~51071 (Remillard et al.\ 1998).
\xte1550 flared to 6.8~Crab on MJD~51075--51076, making it the 
  brightest X-ray nova observed with RXTE to date (Remillard et al.\ 1998).

 On MJD~51065, an optical counterpart with a magnitude of V=16 
  was identified (Orosz, Bailyn \& Jain 1998), 
  and a radio counterpart was discovered at the optical position with a 
  flux density of 10$\pm$2.5~mJy (Campbell-Wilson et al.\ 1998).
An orbital period of $\sim$1.54 days was attributed to the system 
  (Jain et al.\ 2001).
Orosz et al.\ (2002) have found dynamical evidence for 
  a black hole in \xte1550. Based on optical spectra
  of the companion star  (the dereddened colour
  suggests spectral type G/K), they find that the most
  likely value of the mass of the compact object lies in the 
  range $8.63\Msun \leq M_1 \leq 11.58\Msun$, which is well above
  the maximum mass of a stable neutron star.  

 \xte1550 showed a relativistic ejection event in the radio wavebands
   during the 1998 outburst (Hannikainen et al.\ 2001).
A gradual deceleration in the relativistic jets was inferred from Australia Telescope
  Compact Array (ATCA) observations in 2000 and 2002 (Corbel et al.\ 2002).
The radio data are consistent with a synchrotron origin for the emission
   from the jets.
The jet components of \xte1550 were seen also  in {\it Chandra} images  
   (Corbel et al.\ 2002; Kaaret et al.\ 2003; Tomsick et al.\ 2003).  
Interestingly, the extrapolation of the radio spectrum
  to the X-ray wavelengths is consistent with the X-ray flux density
  derived from the {\it Chandra} observations at similar epochs. 
If the X-ray emission is direct synchrotron emission
  from the same population of relativistic electrons
  that gives rise to the radio emission,
  the energies of these  electrons must reach $\sim$10~TeV (Corbel et al.\ 2002).
This scenario requires a very efficient acceleration mechanism  operating in \xte1550  
  so as to populate the electrons in the very high-energy ranges.

Timing analysis showed evolving X-ray QPOs (Cui et al.\ 1999; Titarchuk \& Shrader 2002) 
  and hard lags (Wijnands, Homan \& van der Klis 1999),  
   similar to those often seen in other  black hole binaries.  
The source also showed non-trivial X-ray spectral variability, spectral state  transitions 
   and probably hysteresis during the 1998 and 2000 outbursts 
  (Corbel et al.\ 2001; Homan et al.\ 2001; 
  Kubota \& Done 2004; Wu, Liu \& Li 2007; Xue, Wu \& Cui 2008). 
In the  2--30~keV range, the spectra  could generally be fitted by a three-component model,  
  consisting of a thermal (multi-colour) disc, a  power law  
  and a 6.4~keV Gaussian line component (Sobczak et al.\ 1999; Kubota \& Done 2004), 
  but the relative strengths of the components changed as the outburst proceeded. 
 The origin of the X-rays from \xte1550 is the subject of much debate.
Some suggest that the X-rays (at least the large-scale jet components) 
  are direct or Comptonized synchrotron emission 
  (see e.g.\ Tomsick et al.\ 2003; Wang, Dai \& Lu 2003; Corbel, Tomsick \& Kaaret 2006).
On the other hand, for the X-ray bright case at least,     
  a recent study based on {\it Chandra} and {\it RXTE} data
  argues that the keV X-rays are mainly emission arising from the accretion flows 
  (Xue, Wu \& Cui 2008).

 The 1998 outburst of \xte1550 can be divided into five phases
   based on the morphology of the X-ray light curve observed by {\it RXTE}/ASM.
The phases are characterised by
  (i) a fast rise in the X-ray flux, (ii) a slower rise in the X-ray flux,
  (iii) a giant flare, (iv) a period of stand-still in the flux
  and (iv) an exponential-like decay (Wu et al.\ 2002).
At the onset of the outburst,
  the initial rise in the hard (20--100~keV) X-ray flux was impulsive,
  which can be seen in the BATSE data.
The rise in the soft ($<$20~keV) X-ray flux was, in contrast, less rapid.
The difference in the rise times of the soft and hard X-rays
  can be attributed to the accreting matter
  having two different distributions of angular momenta
  (see Beloborodov \& Illarionov 2001; Wu et al.\ 2002; Smith, Heindl \& Swank 2002).
While matter with low angular momentum
  surges into the black hole on a free-fall timescale,
  matter with higher angular momentum diffuses inward
  forming a viscous accretion disk.
A possible scenario for the mass transfer in \xte1550 was proposed by Wu et al.\ (2002), 
  which explains  the co-existence of material with high and low angular momenta
  that gives rise to the respective soft and hard X-ray bursts.   

Here we present  radio observations of \xte1550 during the 1998 September outburst. 
The paper is organized as follows. 
In \S~\ref{radiobs} we show the technical details and the results of the radio observations;
in \S~\ref{compare} we compare the radio outburst behaviour of \xte1550 
  with that of another galactic superluminal source, GRO~J1655$-$40. 
A summary is given in \S~\ref{summary}. 

\section{Radio observations}
\label{radiobs}

The initial radio detection of \xte1550 was made on MJD~51065 
  with the Molonglo Observatory Synthesis Telescope (MOST) 
  of the University of Sydney.
The MOST observed the source on twelve occasions
  at 843~MHz between MJD~51065 and 51092,
  while the  ATCA observed it at 1.4, 2.3, 4.8 and 8.6~GHz between
  MJD~51073 and 51085.
Very Long Baseline Interferometry (VLBI) 
  images were obtained with the Australian Long Baseline Array (LBA)
  at 2.29 and 8.4~GHz about fifteen days after the first detection
  (MJD~51080--51082). 

\subsection{MOST and ATCA}
\label{moat}

The MOST operates on the principle of earth-rotation aperture
  synthesis at a frequency of 843~MHz. 
It is a 1.6~km East-West array,
  consisting of two cylindrical paraboloids of dimension 778m$\times$12m 
  separated by a gap of 15m, giving a resolution of
  43$^{''}\times$43$^{''} {\rm cosec} |\delta|$ (Mills 1981; Robertson 1991). 

\begin{figure}
\includegraphics[width=8.0cm,angle=0]{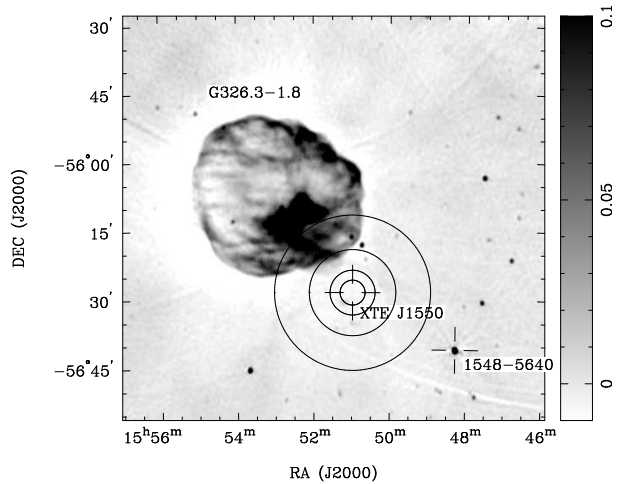}   
\caption{The MOST image at 843~MHz, showing the ATCA fields of view
   at 1.4, 2.3, 4.8 and 8.6~GHz at the position of \xte1550
   and the supernova remnant G326.3$-$1.8. To the lower right
   is SUMSS~J154816$-$564057, used as the reference source for estimating
   the distance to \xte1550 (see Sec.~\ref{hi}). } 
   \label{fig-most}
\end{figure}

Following the initial detection of \xte1550, the source was monitored 
  over the following 27~days resulting in 12 observations, 
  using partial synthesis observations 
  with integration times ranging from 1 to 9~hours. 
The calibration sources used were PKS~B1421$-$490 and B1934$-$638. 
Flux density measurements were complicated by sidelobes from the
  nearby radio-bright supernova remnant G326.3$-$1.8 (Whiteoak \& Green 1996),
  clearly seen in Fig.~\ref{fig-most} which shows the
  image of the field of \xte1550 at 843~MHz.
This necessitated an image differencing approach using a matched 
  reference image of the field obtained earlier in 1998 when 
  \xte1550 was quiescent. 
The flux density estimates are given in Table~\ref{tab-radiobs}, 
  with the errors being the quadrature combination of a 3~mJy rms noise contribution 
  and a 3\% calibration uncertainty. 

\begin{figure*}
\centering   
\includegraphics[angle=0]{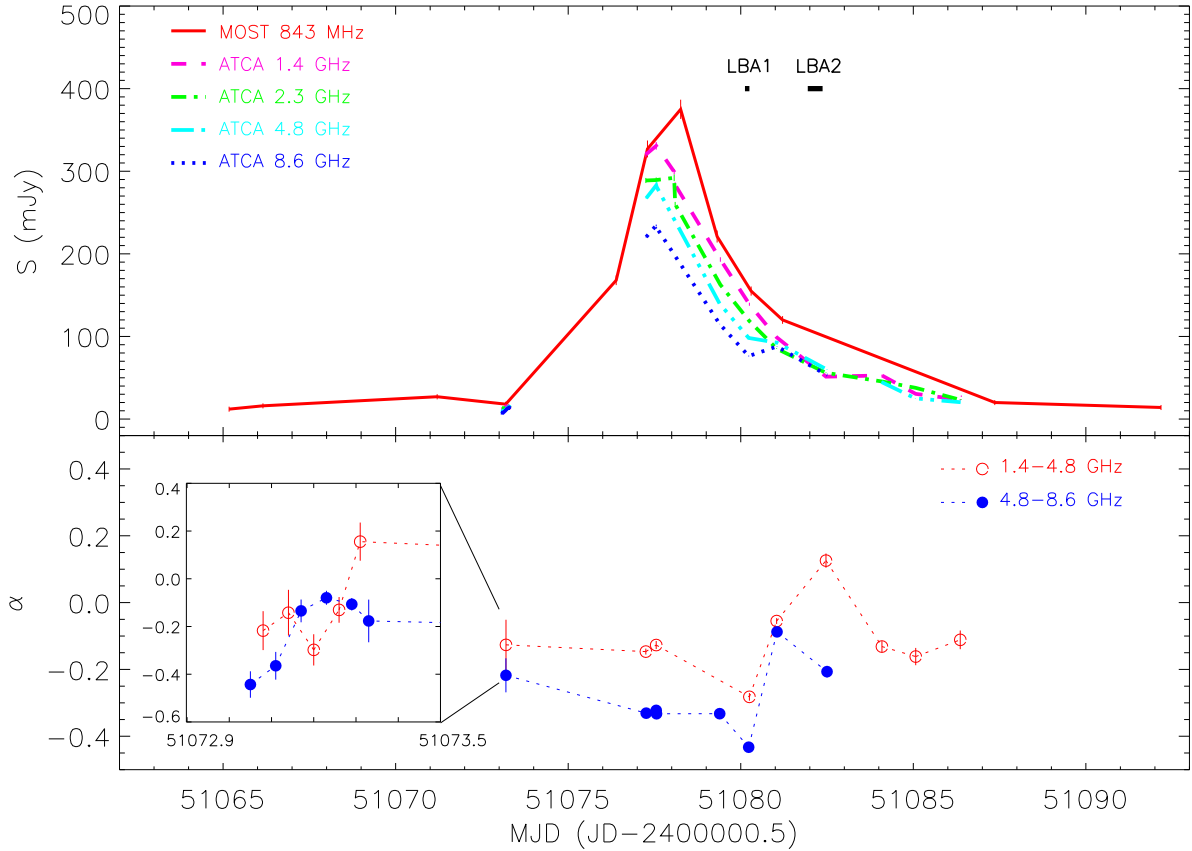}
\caption{The MOST and ATCA  lightcurves
         from 1998 September (top). The epochs (LBA1 corresponds to Epoch 1 and LBA2 to Epoch 2) 
         and durations of the LBA observations are marked.
         The two-point spectral indices
         derived from the ATCA data are in the bottom panel. 
         The data points at $\sim$MJD~51073 are an average of 
         the first five 1.4--4.8~GHz and the first six 4.8--8.6~GHz data points, 
         shown in the inset. The dates for the 1.4~GHz data were used in plotting
         the 1.4--4.8~GHz spectral indices.}
\label{fig-radio}
\end{figure*} 

\begin{table*}
 \centering
  \caption{Flux densities of \xte1550 at all five observing frequencies. The MJD refers to mid-observation.}
  \label{tab-radiobs} 
   \begin{tabular}{crccrrrr}\hline
     \multicolumn{2}{c}{MOST} \vline & \multicolumn{6}{c}{ATCA}  \\ \hline
    MJD   & S$_{0.843}$ &   MJD    & S$_{1.4}$     & S$_{2.3}$     &    MJD   &  S$_{4.8}$    & S$_{8.6}$    \\
          &    (mJy)    &          & (mJy)        &   (mJy)       &          &   (mJy)      &     (mJy)    \\ \hline
 51065.18 & 12$\pm$3    &          &               &               &          &               &               \\
 51066.16 & 16$\pm$3    &          &               &               &          &               &               \\ 
 51071.20 & 27$\pm$3    &          &               &               &          &               &               \\ 
 51073.19 & 18$\pm$3    & 51073.08 &  13.2$\pm$1.3 &  13.3$\pm$0.4 & 51073.05 &  10.1$\pm$0.2 &   7.8$\pm$0.2 \\ 
          &             & 51073.14 &  11.2$\pm$1.3 &  13.2$\pm$0.4 & 51073.11 &   9.4$\pm$0.2 &   7.6$\pm$0.2 \\  
          &             & 51073.20 &  15.3$\pm$1.2 &  12.1$\pm$0.4 & 51073.17 &  10.6$\pm$0.2 &   9.8$\pm$0.2 \\ 
          &             & 51073.26 &  15.5$\pm$1.0 &  13.4$\pm$0.4 & 51073.23 &  13.2$\pm$0.2 &  12.6$\pm$0.1 \\ 
          &             & 51073.31 &  12.3$\pm$1.2 &  16.6$\pm$0.3 & 51073.29 &  14.9$\pm$0.2 &  14.0$\pm$0.1 \\
          &             &          &               &               & 51073.33 &  14.3$\pm$0.5 &  12.9$\pm$0.5 \\
 51076.39 & 168$\pm$6   &          &               &               &          &               &               \\ 
 51077.30 & 327$\pm$10  & 51077.25 & 320.2$\pm$3.4 & 288.8$\pm$3.1 & 51077.26 & 267.3$\pm$0.7 & 220.4$\pm$0.7 \\
          &             & 51077.55 & 330.8$\pm$4.7 & 289.4$\pm$3.1 & 51077.55 & 282.9$\pm$1.2 & 234.4$\pm$1.0 \\
          &             &          &               &               & 51077.56 & 281.3$\pm$1.2 & 231.7$\pm$1.0 \\
 51078.26 & 375$\pm$12  & 51078.07 & 299.4$\pm$3.2 & 292.1$\pm$4.0 &          &               &               \\ 
          &             & 51078.10 & 283.5$\pm$3.2 & 260.1$\pm$2.8 &          &               &               \\
 51079.32 & 221$\pm$7   & 51079.41 & 193.2$\pm$2.7 & 162.6$\pm$0.7 &          &               &               \\
          &             &          &               &               & 51079.39 & 139.6$\pm$0.2 & 115.0$\pm$0.2 \\
 51080.30 & 155$\pm$6   & 51080.25 & 139.0$\pm$1.9 & 118.6$\pm$0.6 & 51080.23 &  98.2$\pm$0.3 &  76.3$\pm$0.5 \\
 51081.21 & 120$\pm$5   & 51081.04 &  99.0$\pm$1.0 &  85.3$\pm$0.5 & 51081.05 &  92.5$\pm$0.3 &  87.9$\pm$0.3 \\
          &             & 51082.47 &  51.3$\pm$1.3 &  56.0$\pm$0.5 & 51082.50 &  59.9$\pm$0.4 &  53.1$\pm$0.4 \\
          &             & 51084.09 &  52.8$\pm$1.0 &  45.5$\pm$0.5 & 51084.08 &  44.9$\pm$0.3 &               \\
          &             & 51085.07 &  30.5$\pm$0.9 &  37.8$\pm$0.4 & 51085.06 &  25.0$\pm$0.3 &               \\
          &             & 51086.36 &  23.4$\pm$0.8 &  23.3$\pm$0.3 & 51086.37 &  20.4$\pm$0.1 &               \\
          &             & 51086.39 &  27.7$\pm$0.9 &  23.7$\pm$0.4 &          &               &               \\  
 51087.36 & 20$\pm$3    &          &               &               &          &               &               \\ 
 51092.18 & 14$\pm$3    &          &               &               &          &               &               \\ 
          &             & 51109.23 &   6.3$\pm$0.7 &   3.2$\pm$0.3 &          &               &               \\  \hline

   \end{tabular}
\end{table*}  

The ATCA is an earth-rotation aperture synthesis array consisting
  of six 22-m antennas on a 6-km baseline, 
  operating at that time at frequencies of 1.2--10.2~GHz 
  with bandwidths of 128~MHz in two linear polarizations. 
The amplitude and bandpass calibrator was PKS~B1934$-$638, 
  while the phase calibrators were B1549$-$790 at 1.4 and 2.3~GHz, 
  and B1554$-$64 at 4.8 and 8.6~GHz.

The top panel of Figure~\ref{fig-radio} shows the MOST 843~MHz and 
  ATCA lightcurves at four observing frequencies.
The MOST 843~MHz lightcurve is reproduced here from Wu et al. (2002).
Also plotted are the global two-point spectral indices calculated
  from the ATCA data (Fig.~2 bottom). 
All the radio data, including observing dates, flux densities at all
  five observing frequencies, and global two-point spectral indices
  are summarized in Tables~\ref{tab-radiobs} and \ref{tab-specind}.
 
 Following the initial detection of \xte1550 on MJD~51065, the
  MOST monitored the source at 843~MHz over the next 27 days
  (Figure~\ref{fig-radio}). 
Between MJD~51065 and 51073, the source flux density remained
  between 10 and 30~mJy. 
After MJD~51073 the flux density began to increase, reaching 
  168~mJy on MJD~51076, and peaking on MJD~51078 with 
  a flux density of 375~mJy. 
The flux density then declined to 120~mJy over the next
  three days, with a continuing decline to 
  14~mJy on MJD~51092, after which the monitoring ceased.

The ATCA started observing \xte1550 on MJD~51073 while the
  flux density was still  low, $\sim8-16$~mJy at all frequencies
  (Figure~\ref{fig-radio}). 
After the rise to 168~mJy detected with the MOST, the ATCA resumed
  observing the source and monitored it over the next ten days.
The ATCA flux densities range from $330.8$~mJy
  at 1.4~GHz to $234.4$~mJy at 8.6~GHz at the peak of the lightcurves, 
  after which they then followed the same decline as the MOST,
  reaching a level of around 20~mJy at 4.8~GHz on MJD~51086.5 
  when monitoring ceased.

The  bottom panel of Figure~\ref{fig-radio} shows the evolution of
  two-point spectral indices $\alpha$ ($S_{\nu} \propto \nu^{\alpha}$)
  derived from the 1.4--4.8~GHz ($\alpha_{1.4-4.8}$) 
  and the 4.8--8.6~GHz ($\alpha_{4.8-8.6}$) data. 
$\alpha_{1.4-4.8}$ varies between $-0.28$ and $-0.12$ 
  and $\alpha_{4.8-8.6}$ between $-0.43$ and $-0.09$ (Table~\ref{tab-specind}). 
The inset in the bottom panel
  shows the evolution of 
  $\alpha_{1.4-4.8}$ and $\alpha_{4.8-8.6}$ in detail -- for clarity in the main plot
  the first five points were averaged for $\alpha_{1.4-4.8}$ and the first six points 
  for $\alpha_{4.8-8.6}$.
 
\begin{table*}
 \centering
  \caption{Spectral indices. The MJD refers to mid-observation.}
  \label{tab-specind} 
   \begin{tabular}{crcrcr}\hline
    MJD   & $\alpha_{0.843-8.6}$   &   MJD   & $\alpha_{1.4-4.8}$ &  MJD  &  $\alpha_{4.8-8.6} $  \\ \hline
          &                 & 51073.08 & $-$0.22$\pm$0.08 & 51073.05 & $-$0.44$\pm$0.06 \\
          &                 & 51073.14 & $-$0.14$\pm$0.10 & 51073.11 & $-$0.36$\pm$0.06 \\  
 51073.19 &  $-$0.22$\pm$0.06 & 51073.20 & $-$0.30$\pm$0.07 & 51073.17 & $-$0.13$\pm$0.05 \\ 
          &                 & 51073.26 & $-$0.13$\pm$0.05 & 51073.23 & $-$0.08$\pm$0.03 \\ 
          &                 & 51073.31 & 0.16$\pm$0.08 & 51073.29 & $-$0.11$\pm$0.03 \\
          &                 &          &                & 51073.33 &  $-$0.18$\pm$0.09 \\
          &                 & 51077.25 & $-$0.15$\pm$0.01 & 51077.26 & $-$0.33$\pm$0.01 \\
 51077.30 & $-$0.17$\pm$0.02  & 51077.55 & $-$0.13$\pm$0.01 & 51077.55 & $-$0.32$\pm$0.01 \\
                    &                                   &                   &                                    & 51077.56 & $-$0.33$\pm$0.01 \\ 
 51078.26 & $-$0.30$\pm$0.01  &          &                &          &                \\ 
 51079.32 & $-$0.26$\pm$0.01  & 51079.41 & $-$0.26$\pm$0.01 & 51079.39 & $-$0.33$\pm$0.01 \\
          &                 &          &                &          &                \\
 51080.30 & $-$0.30$\pm$0.01   & 51080.25 & $-$0.28$\pm$0.01 & 51080.23 & $-$0.43$\pm$0.01 \\
 51081.21 & $-$0.15$\pm$0.02  & 51081.04 & $-$0.06$\pm$0.01 & 51081.05 & $-$0.09$\pm$0.01 \\
          &                 & 51082.47 &  0.12$\pm$0.02 & 51082.50 & $-$0.21$\pm$0.02 \\
          &                 & 51084.09 & $-$0.13$\pm$0.02 &          &                \\
          &                 & 51085.07 & $-$0.16$\pm$0.03 &          &                \\
          &                 & 51086.36 & $-$0.11$\pm$0.03 &          &                \\ \hline  
   \end{tabular}
\end{table*} 

\subsection{Distance estimate using the H{\sc i} spectrum}
\label{hi}

\begin{figure*}
\centering   
\includegraphics[width=16cm]{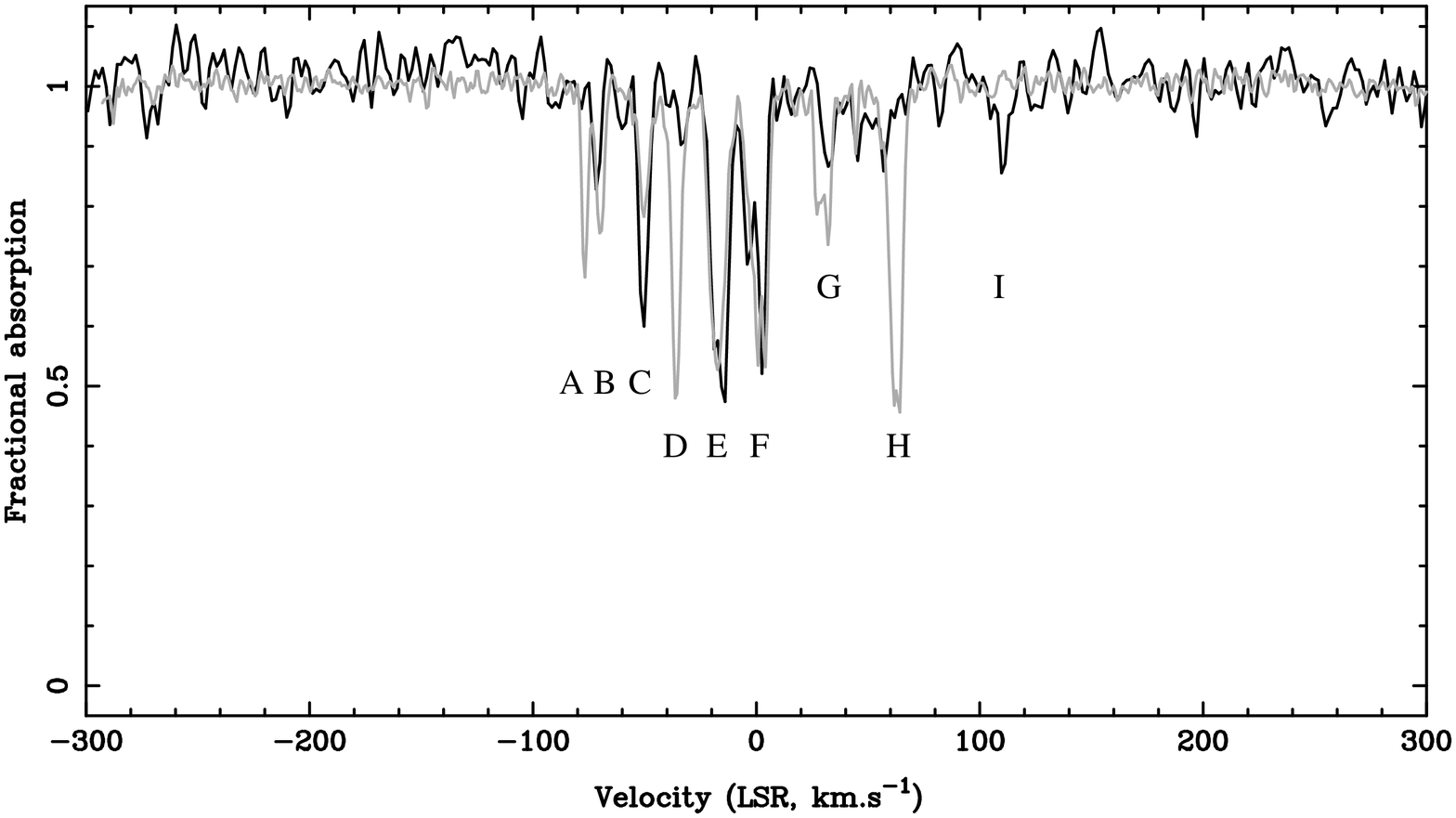}   
\caption{The H{\sc i} spectrum of \xte1550 represented in black
and a spectrum taken towards SUMSS~J154816$-$564057 in grey.
Absorption of the continuum emission from \xte1550 is seen clearly in 
the features labelled C, E \& F.}
\label{fig-hi}
\end{figure*}

\subsubsection{Observing details and data reduction}

On 1998 Sep 22 we measured H{\sc i} absorption towards \xte1550.
We employed 3 antennas of the ATCA, and set the correlator to produce 1024 channels 
  over a bandwidth of 4~MHz centred on 1420~MHz.
Two linear polarisation products were recorded.

To provide an extragalactic reference, 
   on 2000 Apr 5 and 10 we observed the source SUMSS~J154816$-$564057 (see Fig.~1)
   using all 6 ATCA antennas and the same correlator setting.
SUMSS~J154816$-$564057 is 25 arcmin south-west of \xte1550 at
   RA=15h48m16.5s, DEC=$-$56$^{\circ}40^{\prime}57^{\prime\prime}$ (J2000), 
   and its flux density is 0.78$\pm$0.04~Jy at 843~MHz.
In both cases the phase centre of the interferometer 
   was positioned 1~arcminute to the South of the target source.

The data were flagged and calibrated with \textsc{miriad},
   following standard procedures.
During reduction, the data were Hanning-smoothed 
   and every other channel was discarded. 
The 512 channels in the resulting spectrum are separated by 1.65~\kms.

The spectra were obtained by shifting the phase centre of the array to the source position, 
  and vector-averaging the Stokes $I$ visibilities (\textsc{miriad} task \textsc{uvspec}). 
After several experiments with different choices of $uv$-data, 
  baselines shorter than 2~k$\lambda$ were excluded, 
  in order to avoid artifacts caused by the nearby supernova remnant G326.3$-$1.8. 
Both scans of SUMSS~J154816$-$564057 were averaged together, 
  after intercomparison showed negligible differences between them. 
The continuum flux density of \xte1550 was 0.280$\pm0.005$~Jy 
  and that of SUMSS~J154816$-$564057 was 0.465$\pm0.002$~Jy;
  the net on-source integration times (after flagging and $uv$-selection) 
  were 60 and 132 minutes, respectively.

\subsubsection{Interpretation}
\label{interp}

Figure~\ref{fig-hi} shows the two H{\sc i} absorption spectra
superimposed: \xte1550 in black and SUMSS~J154816$-$564057 in grey. These
spectra have been further Hanning-smoothed over 3 channels, with all
channels retained. Several features have been identified, and listed in
Table~\ref{tab-hi}. The estimates of the Galactocentric radius and
heliocentric distance to the clouds producing the absorptions are based on
the Galactic rotation curve given by Fich, Blitz \& Stark (1989).

Comparison of these spectra is complicated by significant differences in
signal to noise. In the normalised units of Figure~\ref{fig-hi} we
estimate the rms noise to be 0.04 in the \xte1550 spectrum and 0.013 in
the SUMSS~J154816$-$564057 spectrum. In the spectrum toward
SUMSS~J154816$-$56405, we are confident that all the features identified
by letters A to H in Figure~\ref{fig-hi} and Table~\ref{tab-hi} are
absorptions produced by H{\sc i} clouds along the line of sight, and not
instrumental artefacts.  In the spectrum toward \xte1550 we have no
hesitation in identifying the features near C, E and F as real aborptions,
while the feature near B is marginally significant. However, we do not
consider feature I or the feature near G to be strong enough to be
considered real absorptions, given the noise level in the spectrum.
Features D and H are deep enough that, should the sightline toward
\xte1550 intersect gas from the same cloud under the same excitation
conditions, we should have detected these absorptions above the noise.

The three clear H{\sc i} absorptions found in the spectrum of \xte1550
match well with corresponding features seen in SUMSS~J154816$-$564057. The
velocity of feature F in \xte1550 is uncertain but the overall profile
match with SUMSS~J154816$-$564057 is good so we have adopted the same
V$_{\rm lsr}$.

The observed absorptions allow us to constrain the distance to \xte1550.
The strongest constraint comes from feature C, which implies a distance
greater than 3.3~kpc. However the velocity of this feature is associated
with two possible distances (as are all the features A--F). We believe
that the far distance (10.6~kpc) for feature C is effectively ruled out by
the absence of feature A (4.88~kpc).  If \xte1550 lies beyond 3.3~kpc, our
non-detection of feature D (2.52~kpc) requires either that this sightline
misses the cloud seen against SUMSS~J154816$-$564057 or that there is a
dearth of H{\sc i} along this part of the sightline. Emission spectra at
the position of \xte1550 (McClure-Griffiths et al.~2005) show emission
peaks at velocities matching the strong absorptions C, E and F, but weaker
emission near $-$30~km\,s$^{-1}$ suggesting there could indeed be a dearth
of H{\sc i} at the velocity corresponding to D. In setting an upper limit
for the distance to \xte1550 we prefer to use the absence of a counterpart
to feature A (4.88~kpc) rather than argue over the reality of the feature
near B (4.45~kpc).

In summary, we interpret the H{\sc i} spectra as showing the distance to
\xte1550 is probably in the range 3.3--4.9~kpc. This puts the source
slightly farther away than the estimate of 2.5~kpc based on the equivalent
width of the NaID absorption line in the optical spectrum
(Sanchez-Fernandez et al.~1999).

\begin{table}
\caption{Features in the H{\sc i} spectra of \xte1550 and
 SUMSS~J154816$-$564057.}
\label{tab-hi}
\begin{tabular}{rrrl} \hline
V$_{\rm lsr}$ (km/s) & R$_{\rm gal}$ (kpc) & Distance (kpc)& \\ \hline
\multicolumn{3}{l}{SUMSS~J154816$-$564057} & \\ \hline
64.3 & 17.37 & 23.65 & H \\
31.8 & 11.28 & 17.27 & G \\
27.4 & 10.77 & 16.7 & G \\
0.7 & 8.47 & 0.04 & F \\
$-16.2$ & 7.46 & 1.3 & E \\
$-35.4$ & 6.57 & 2.52 & D \\
$-50.1$ & 6.02 & 3.37 & C \\
$-69.2$ & 5.43 & 4.45 & B \\
$-76.3$ & 5.24 & 4.88 & A \\ \hline
\multicolumn{3}{l}{\xte1550} & Id \\ \hline
110.4 & 69.9 & 76.8 & I? \\
0.7: & 8.46 & 0.04 & F \\
$-12.7$ & 7.65 & 1.06 & E \\
$-49.2$ & 6.05 & 3.32 & C \\ \hline
\end{tabular}
\end{table}

\subsubsection{Possible association with G326.3$-$1.8}

The overlap in distance between XTE J1550$-$564 (4.1 $\pm$ 0.8 kpc)
   and G326.3$-$1.8 (4.1 $\pm$ 0.7 kpc; Rosado et al.~1996) and their proximity
   in the plane of the sky raises the interesting question of a possible
   connection.  
Given the estimated age of the supernova remnant of $1.04 \pm
   0.09 \times 10^4$ yr (Kassim, Hertz \& Weiler 1993), we considered the possibility that
   \xte1550 was ejected during the supernova explosion.  
The implied transverse velocity, however, is $\sim 3000$ km\,s$^{-1}$ which is
   implausibly high.  
We therefore conclude that \xte1550 and G326.3$-$1.8 are not associated. 
Further evidence comes from {\it Chandra} and {\it XMM-Newton} observations 
  that may have identified the pulsar and associated X-ray synchrotron nebula close 
  to the radio plerion (Plucinsky et al.\ 2004).

\subsection{LBA images}
\label{lba}

\begin{figure}
\centering   
\includegraphics[width=8.7cm,angle=0]{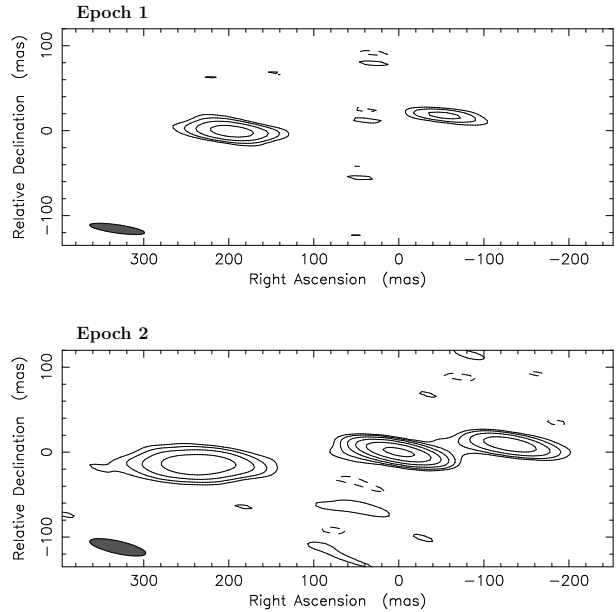}
\caption{LBA images at 2.29~GHz from 1998 Sep 24 (Epoch 1, top) and 1998 Sep 25/26
(Epoch 2, bottom). The beam is shown in the lower left-hand corner. The zero
coordinate is arbitrary.  Reproduced from Hannikainen et al.\ (2001).}
\label{fig-vlbi}
\end{figure}

The LBA observations were carried out on 1998 Sep 24 (Epoch 1)
   and 1998 Sep 25/26 (Epoch 2) and are summarized in Table~\ref{tab-vlbiobs}. 
Preliminary results and their interpretation were reported in 
   Hannikainen et al.~(2001).
Note, however, that the mid-point MJDs  quoted in that paper are erroneous,
   and we report the correct dates and times here. 
Figure~\ref{fig-vlbi} shows the two LBA images at 2.29~GHz and
   Table~\ref{tabfd} summarizes the flux densities of the components.
The  zero coordinate in the images is arbitrary, as phase-referenced 
    observations were not performed on either day.

\begin{figure}
\centering   
\includegraphics[width=8cm,angle=0]{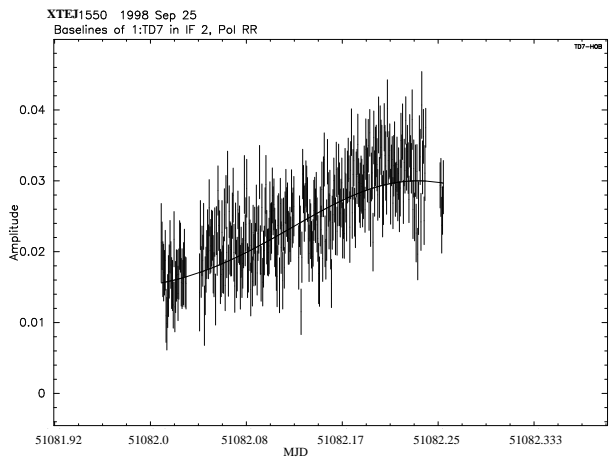}
\caption{The observed visibility amplitudes on the DSS43-Hobart baseline at 
8.4~GHz on 1998 Sep 25 (MJD~51082) and a fit of a simple source model
consisting of a single elliptical Gaussian component (curve). }
\label{fig-gauss}
\end{figure}

\begin{table}
\caption{LBA observing log}
\label{tab-vlbiobs}
\begin{tabular}{lllll} \hline
Date   &   T$_{\rm start}$ & T$_{\rm end}$ & Antenna & Frequency \\
       &     (UT)           & (UT)          &         & (GHz)   \\ \hline
Epoch 1       &            &                  &                        &   \\
\,\,\,\,\,1998 Sep 24 & 02:50          & 06:00         & DSS45 (34m) & 2.29 \& 8.4\\
\,\,\,\,\,Mid-point: & \multicolumn{2}{l}{51080.18} & Hobart (26m) & 2.29 \& 8.4 \\             
\,\,\,\,\,(MJD)  &                      &               & Mopra (22m)  & 2.29   \\  

       &                     &               &              & \\
Epoch 2  &  &  &  & \\
\,\,\,\,\,1998 Sep 25 & 22:30         & 09:00         & DSS43 (70m) &  2.29 \& 8.4\\
\,\,\,\,\,Mid-point: & \multicolumn{2}{l}{51082.15}& Hobart (26m)&  2.29 \& 8.4\\           
\,\,\,\,\,(MJD)  &                      &               & Mopra (22m)    &  2.29 \\           
       &                      &               & ATCA  & 2.29  \\ 
       &                      &               & ~~~~~~(6$\times$22m) & \\ \hline
\end{tabular}
\end{table}
 
\begin{table*}
\caption{Component positions and flux densities at 2.29~GHz. The uncertainties in the flux densities are of the order 10--20\%.}
\label{tabfd}
\begin{tabular}{lccc|c|ccc} \hline
Date & \multicolumn{2}{c}{Relative position (mas)} & 
  Position angle &\multicolumn{3}{c}{Flux density (mJy)} & Total \\ 
 & East & West & (degrees) & East & Central & West & (mJy) \\ \hline 
Epoch 1 & 200* & 55* & $-86.1^{+1.5}_{-0.9}$ & 71 & --- & 20 & 91 \\ 
Epoch 2 & $254^{+15}_{-12}$ & $130\pm 12$ & $-86.1\pm 0.8$ & 19 & 25 & 8 & 52 \\ \hline
\multicolumn{8}{l}{*Positions measured relative to an arbitrary origin.  
Component separation is $255^{+15}_{-19}$ mas.}
\end{tabular}
\end{table*} 

An additional component is seen in the image from Epoch 2 compared
   to Epoch 1, but as the image centre is arbitrary, it 
   is not clear how the components are related. 
We note, however, that the ATCA two-point spectral indices start to flatten together right after
   the first LBA observation (Figure~\ref{fig-radio}).
Shortly thereafter, the 4.8--8.6~GHz spectral index steepens again slightly, while the
   1.4--4.8~GHz spectral index continues to flatten.
The latter starts to steepen only some time during or after the second LBA observation. 

We infer, therefore, that a new outburst has occurred in \xte1550 between Epoch 1
   and Epoch 2 and associate the central component in the bottom panel of Figure~\ref{fig-vlbi}
   with the flaring core and the two outer components with the ejecta seen in the top panel.
Single baseline 8.4~GHz data from Epoch 2 (Fig.~\ref{fig-gauss}) show a new unresolved 
  component not present in the data from Epoch 1, which  
  supports this interpretation.

The increase in 8.4~GHz visibility amplitude on the DSS43-Hobart baseline 
   peaked at MJD~51082.21 (Fig.~\ref{fig-gauss}).
Such a change may be due to structure in the source and/or an intrinsic 
  change in the source over the period of the observation. 
The peak in the visibility amplitude on this baseline corresponds to a 
  position angle in the $(u,v)$ plane of $6.5\pm3$~degrees which, for a simple
  elongated source, corresponds to a position angle of the major
  axis in the image plane of $-83.5\pm3$~degrees, consistent 
  with the position angles of the outer components at 2.29~GHz. 
Such a coincidence provides strong evidence 
  that the variation in visibility can be explained by 
  a resolved component at 8.4~GHz. 
Simple model fits to the data place an upper limit on the angular size of this
  component of $\sim5$~mas. 
This unambiguously identifies the central source as the core.  

Table~\ref{tabfd} shows the relative positions of the components in
Fig.~\ref{fig-vlbi} at both epochs. For Epoch 1, the position of
the weaker component (West) was measured with respect to the stronger
component (East), while for Epoch 2 the positions of the outer
components were measured with respect to the central component. The
errors were estimated using {\sc difwrap} (Lovell 2000). 
The table also shows the position angles defined by the outer components. 
We note that Corbel et al.\ (2002) have found the
same position angles from their observations in 2000 and 2002. \\

\begin{figure}
\centering   
\includegraphics[width=8cm]{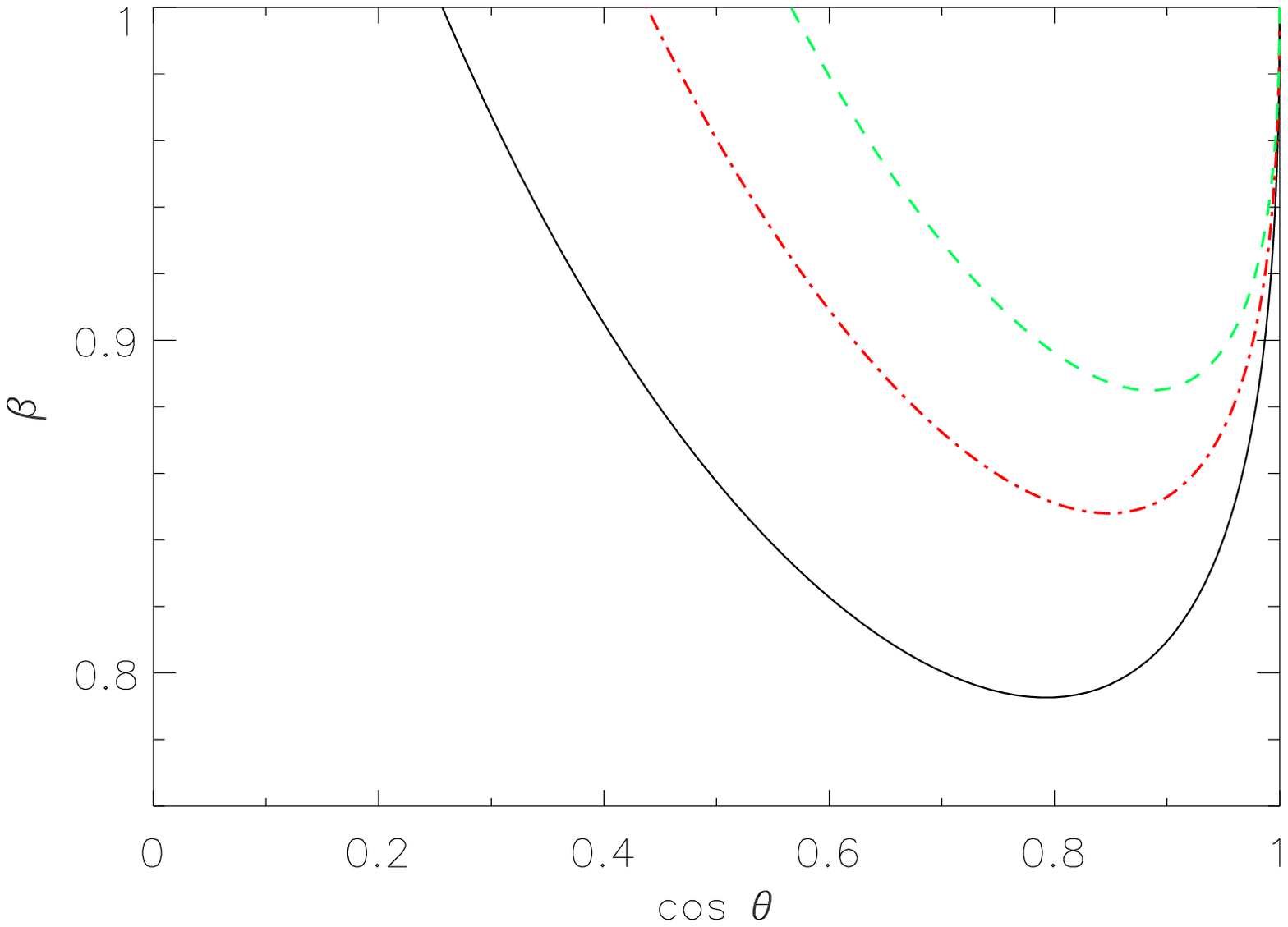} 
\includegraphics[width=8cm]{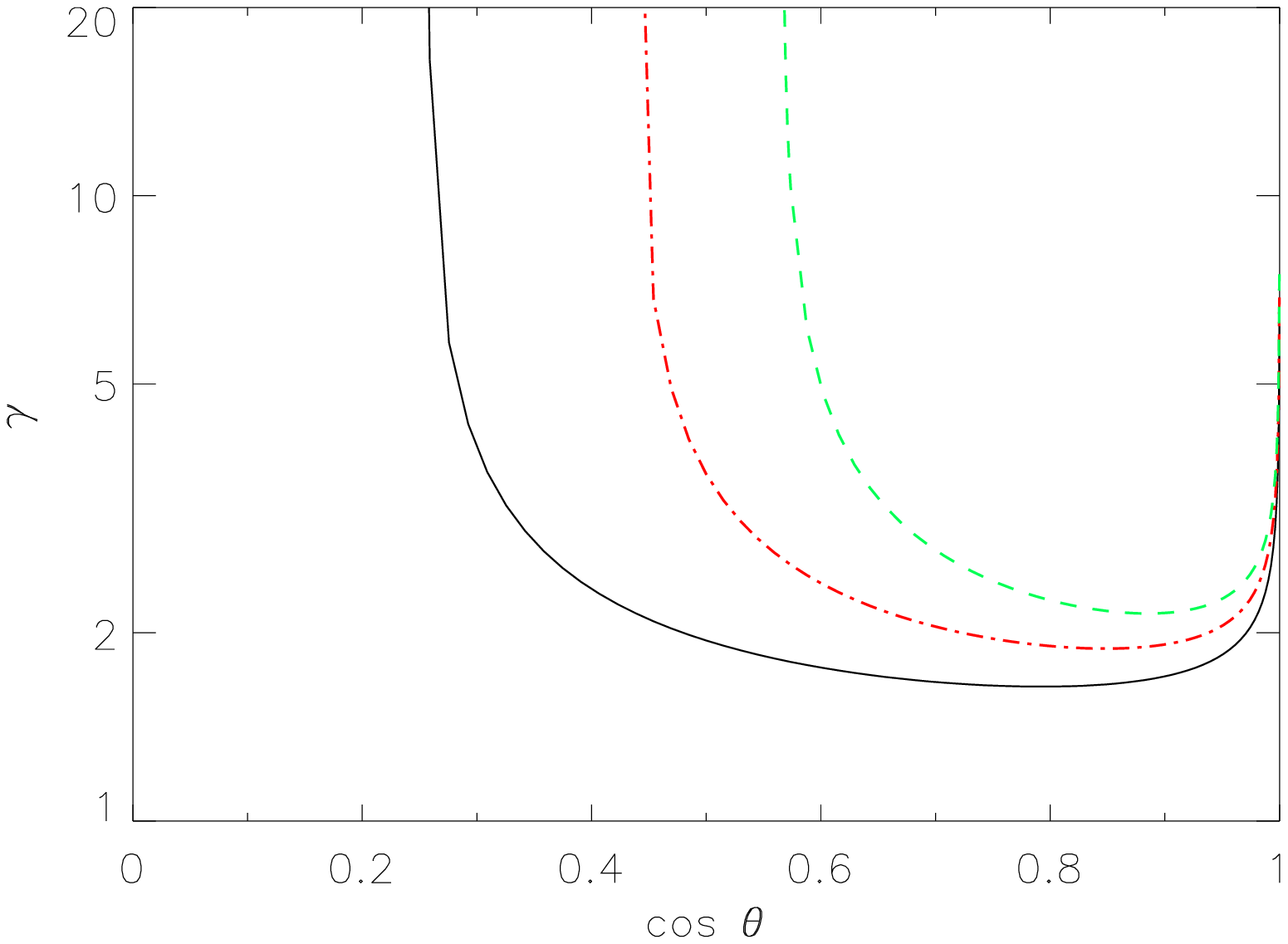}   
\caption{The top panel shows the inferred separation velocity of the ejecta,
$\beta$, for various viewing angles, $\theta$, and 
the bottom panel shows the corresponding $\gamma$
for $v=1.3c$ (solid black), 1.6$c$ (dot-dash red), and 1.9$c$ (dash green).}
\label{fig-beta}
\end{figure}

On the basis of the above interpretation, the ejecta have moved apart by
   approximately 130~mas in $\sim2$ days.
At the closer distance of 3.3~kpc, this implies an apparent
  separation velocity of $1.3c$ for the ejecta, while for 
the farther distance of 4.9~kpc the separation velocity is $1.9c$. 

Fig.~\ref{fig-beta} shows the inferred separation velocity of the ejecta, $\beta$,
  and the corresponding $\gamma$ for various viewing angles, $\theta$,
  for three apparent ejecta velocities:  1.3$c$, 1.6$c$, and 1.9$c$.
Thus, $\beta$ is at least greater than $0.8$ and $\gamma$ is 
  at least greater than $1.6$.

\section{XTE~J1550$-$564 and GRO~J1655$-$40}
\label{compare}

 At least five Galactic sources are known to have exhibited apparent superluminal 
  motion: GRS~1915$+$105 (Mirabel \& Rodr{\'\i}guez~1994), 
  GRO~J1655$-$40 (Tingay et al.~1995; Hjellming \& Rupen~1995), 
  XTE~J1748$-$288 (Hjellming et al.~1998), SAX~J1819.3$-$2525 
  (V4641~Sgr; Hjellming et al.~2000) and \xte1550. 
All five sources displayed very different multiwavelength behaviour
  at the time of the ejection outbursts, and in the evolution of the ejecta themselves. 
However, there are several similarities in the multifrequency 
  data for \xte1550 and GRO~J1655$-$40 that we discuss below.

\begin{figure}
\centering   
\includegraphics[width=9cm]{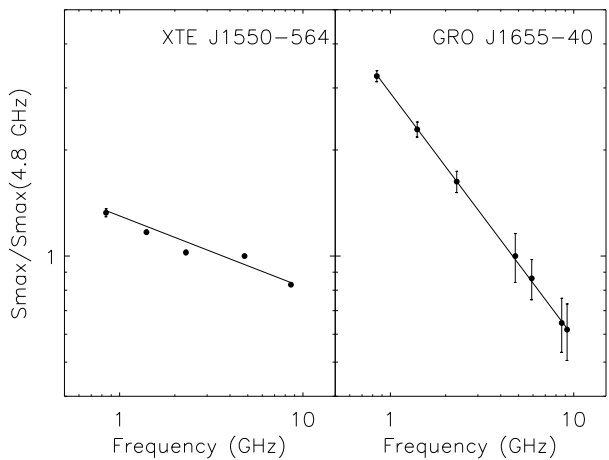}
\caption{Flaring amplitudes for \xte1550 (left) and
   GRO~J1655$-$40 (right; from Stevens et al. 2003). 
   The slopes of the powerlaws are $-0.2$ and $-0.7$ respectively.}
\label{fig-ampl}
\end{figure}

The most obvious similarity between the radio properties of 
  \xte1550 and GRO~J1655$-$40 
is the high expansion velocity of the ejecta, $\geq0.8c$ for \xte1550
  and $\sim0.9c$ for GRO~J1655$-$40 at respective distance estimates
 of $3.3 < d < 4.9$~kpc and $\sim3.2$~kpc.
In addition, the behaviour during the radio outburst is very similar,
  as can be seen when comparing the the flux density lightcurves.
In GRO~J1655$-$40, the lightcurves at all frequencies 
  rise and fall simultaneously, with the lowest frequencies peaking
  with the highest flux densities.
The flux density lightcurves of \xte1550 {\rm appear} to peak 
  simultaneously (Fig.~\ref{fig-radio}, top panel).
Due to lack of simultaneous coverage in the observed 
  frequencies, we
  cannot say with certainty that they peaked at the same time. 
{\rm However}, it is apparent that, as with GRO~J1655$-$40, the 
  flare amplitudes, normalised to 4.8 GHz,
  increase towards lower frequency as shown in Fig.~\ref{fig-ampl},
  albeit with a shallower slope, the powerlaw index being 
  $-0.2$ as opposed to $-0.7$ for GRO~J1655$-$40.

\begin{figure}
\centering   
\includegraphics[width=8cm]{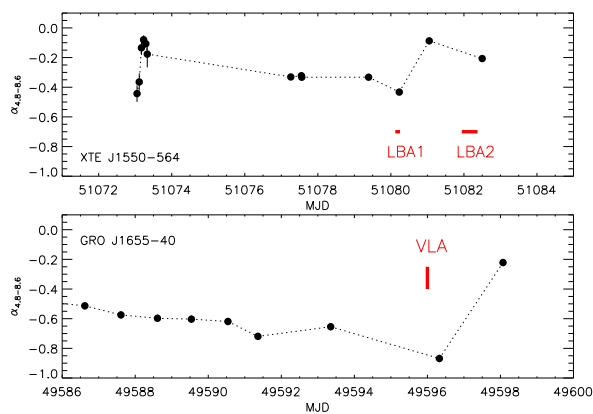}
\caption{Two-point 84.8--8.6~GHz) spectral indices 
   as a function of time for \xte1550 (top) and GRO~J1655$-$40 (bottom). 
   Marked in the top panel are the epochs of the LBA 
   observations shown in Fig.~\ref{fig-vlbi} and, in the bottom panel,  the onset of 
   a major ejection event  in GRO~J1655$-$40 seen with the VLA (Hjellming \& Rupen~1995).
   }
\label{fig-specomp}
\end{figure}

Fig.~\ref{fig-specomp} shows the temporal evolution of two-point
  spectral indices in both \xte1550 and GRO~J1655$-$40 during
  the times of their respective ejection events. 
Although the coverage for GRO~J1655$-$40 was more complete, 
  the figure shows that in both sources, at the time of a
  major recorded ejection event in GRO~J1655$-$40 and during
  the LBA-recorded ejection event in \xte1550, the 4.8--8.6~GHz
  two-point spectral index showed a significant flattening.

\section{Summary}
\label{summary}

We confirm the presence of apparent superluminal motion from \xte1550
   during its 1998 September outburst.  
Comparison between the H{\sc i} absorption spectra towards \xte1550 and the 
   nearby extragalactic source SUMSS~J154816$-$564057 has allowed us to constrain 
   the distance to \xte1550 to lie between 3.3 and 4.9~kpc.  
Based on two-epoch observations with the Australian Long Baseline Array these 
   distances yielded apparent separation velocities of the ejecta of $1.3c$ at the closer 
   distance and $1.9c$ at the farther distance.
After correcting for relativistic effects this leads to an intrinsic ejection velocity $\geq0.8c$.  
The maximum flux density during the 1998 September outburst was 375~mJy at 843~MHz and
  ranged from $\sim330$~mJy to $\sim$234~mJy at 1.4 and 8.6~GHz
  respectively. 
As with the superluminal source GRO~J1655$-$40, the flux densities appear to peak 
   simultaneously at all frequencies, but with a much flatter (but optically thin) spectral index 
   of $-0.2$ compared with $-0.7$ for GRO~J1655$-$40.  
In addition, the spectral evolution during the outburst is reminiscent of that observed in
  GRO~J1655$-$40 during its 1994 outburst.

\section*{Acknowledgements} 
The authors thank Robert Braun and Jim Caswell for helpful discussions 
on the distance to \xte1550. MOST is operated by the University of Sydney 
 and supported by grants from the Australian Research Council.
The Australia Telescope Compact Array and the Mopra Telescope 
  are part of the Australia Telescope 
  which is funded by the Commonwealth of Australia 
  for operation as a National Facility managed by CSIRO. 
We thank the staff of both the University of Tasmania, 
  and the Canberra Deep Space Communications Complex, Tidbinbilla.
DCH acknowledges a Fellowship from the Academy of Finland
and project 212656.


\begin{thebibliography}{99} 
\bibitem[\protect\citeauthoryear{Beloborodov \& Illarionov }{2001}]{beloborodov01} 
         Beloborodov A.M., Illarionov A.E., 2001, MNRAS, 323, 167 
\bibitem[\protect\citeauthoryear{Campbell-Wilson et al. }{1998}]{campbell98} 
         Campbell-Wilson D., McIntyre V., Hunstead R. W., Green A., 1998, IAUC 7010 
\bibitem[\protect\citeauthoryear{Cui et al. }1999]{cui99}
         Cui W., Zhang S.N., Chen W., Morgan E.H., 1999, ApJ, 512, L43 
\bibitem[\protect\citeauthoryear{Corbel et al. }2001]{corbel01}   
         Corbel S. et al., 2001, ApJ, 554, 43         
\bibitem[\protect\citeauthoryear{Corbel et al. }2002]{corbel02}
         Corbel S., Fender R.P., Tzioumis A.K., Tomsick J.A., Orosz J.A.,
         Miller J.M., Wijnands R., Kaaret P., 2002, Science, 298, 196  
\bibitem[\protect\citeauthoryear{Corbel, Tomsick \& Kaaret }2006]{corbel06}
         Corbel S., Tomsick J.A., Kaaret P., 2006, ApJ, 636, 971        
\bibitem[\protect\citeauthoryear{Fich, Blitz \& Stark }1989]{fich89}
          Fich M., Blitz L., Stark A.A., 1989, ApJ, 342, 272
\bibitem[\protect\citeauthoryear{Hannikainen et al. }2000]{hannikainen00}
         Hannikainen D.C., Hunstead R.W., Campbell-Wilson D., Wu K., McKay D.J., 
         Smits D.P., Sault R.J., 2000, ApJ, 540, 521
\bibitem[\protect\citeauthoryear{Hannikainen et al. }2001]{hannikainen01}
         Hannikainen D., Campbell-Wilson D., Hunstead R., McIntyre V., 
         Lovell J., Reynolds J., Tzioumis T., Wu K., 2001, ApSSS, 276, 45
\bibitem[\protect\citeauthoryear{Hjellming \& Rupen }1995]{hjellming95}
         Hjellming R.M., Rupen M.P., 1995, Nature, 375, 464
\bibitem[\protect\citeauthoryear{Hjellming et al. } 1998]{hjellming98}
         Hjellming R.M., Rupen M.P., Mioduszewski A.J., Smith D.A., Harmon B.A., 
         Waltman E.B., Ghigo F.D., Pooley G.G., 1998, AAS, 19310308H
\bibitem[\protect\citeauthoryear{Hjellming et al. }2000]{hjellming00} 
         Hjellming R.M. et al., 2000, ApJ, 544, 977 
\bibitem[\protect\citeauthoryear{Homan et al. }2001]{homan01} 
         Homan J. et al., 2001, ApJS, 132, 377          
\bibitem[\protect\citeauthoryear{Jain et al. }2001]{jain01}
         Jain R.K., Bailyn C.D., Orosz J.A., McClintock J.E., Sobczak G.J., 
         Remillard R.A., 2001, ApJ, 546, 1086
\bibitem[\protect\citeauthoryear{Kaaret et al. }2003]{kaaret03}
         Kaaret P. et al., 2003, ApJ, 582, 945 
\bibitem[\protect\citeauthoryear{Kassim, Hertz \& Weiler }1993]{kassim93}
	Kassim N. E., Hertz P., Weiler K. W., 1993, ApJ, 419, 733
\bibitem[\protect\citeauthoryear{Kubota \& Done }2004]{kubota04}
         Kubota A., Done C., 2004, MNRAS, 353, 980           
\bibitem[\protect\citeauthoryear{Lovell }2000]{lovell00}
         Lovell J., 2000, in ``Astrophysical Phenomena Revealed by Space
         VLBI'', H. Hirabayashi, P.G. Edwards \& D.W. Murphy (eds), 301 
\bibitem[\protect\citeauthoryear{McClure-Griffiths et al. }2005]{SGPS05}
         McClure-Griffiths, N. M., Dickey, J. M., Gaensler, B. M.,
	Green, A. J., Haverkorn, M., Strasser, S. 2005, ApJS, 158, 178
\bibitem[\protect\citeauthoryear{Mills }1981]{mills81}
         Mills B.Y., 1981, PASA, 4, 156
\bibitem[\protect\citeauthoryear{Mirabel \& Rodr{\'\i}guez }1994]{mirabel94}
         Mirabel I.F., Rodr{\'\i}guez L.F., 1994, Nature, 371, 46 
\bibitem[\protect\citeauthoryear{Orosz, Bailyn \& Jain }1998]{orosz98}
         Orosz J.A., Bailyn C.D., Jain R.K., 1998, IAUC 7009 
\bibitem[\protect\citeauthoryear{Orosz et al. }2002]{orosz02}
         Orosz J.A. et al., 2002, ApJ, 568, 845
\bibitem[\protect\citeauthoryear{Plucinsky et al. }2004]{plucinsky04}
         Plucinsky P.P., Dickel J.R., Slane P.O., Gaetz T.J., Sasaki M.,
         Edgar R.J., Smith R.K., 2004, AAS, 20510613P
\bibitem[\protect\citeauthoryear{Remillard et al. }1998]{remillard98}
         Remillard R., Morgan E., McClintock J., Sobczak G., 1998, IAUC 7019  
\bibitem[\protect\citeauthoryear{Robertson }1991]{robertson}
         Robertson J.G., 1991, Aust. J. Phys., 44, 729
\bibitem[\protect\citeauthoryear{Rosado et al. }1996]{rosado96}         
	Rosado M., Ambrocio-Cruz P., Le Coarer E., Marcelin M., 1996, A\&A, 315, 243
\bibitem[\protect\citeauthoryear{Sanchez-Fernandez et al. }1999]{sanchez99}
         Sanchez-Fernandez C. et al., 1999, A\&A, 348, L9
\bibitem[\protect\citeauthoryear{Smith }1998]{smith98}
         Smith D.A., 1998, IAUC 7008 
\bibitem[\protect\citeauthoryear{Smith et al. }2002]{smith02} 
         Smith D.A., Heindl W.A., Swank J.H., 2002, ApJ, 569, 362  
\bibitem[\protect\citeauthoryear{Sobczak et al. }1999]{sobczak99}
         Sobczak G., McClintock J.E., Remillard R.E., Levine A.M., Morgan E.H., 
         Bailyn C.D., Orosz J.A., 1999, ApJ, 517, L121
\bibitem[\protect\citeauthoryear{Stevens et al. }2003]{stevens03}
	Stevens J.A., Hannikainen D.C., Wu K., Hunstead R.W., McKay D., 
	2003, MNRAS, 342, 623 
\bibitem[\protect\citeauthoryear{Titarchuk \& Shrader }2002]{titarchuk02} 
         Titarchuk L., Shrader C.R., 2002, ApJ, 567, 1057  
\bibitem[\protect\citeauthoryear{Tingay et al. }1995]{tingay95}
         Tingay S.J. et al., 1995, Nature, 374, 141 
\bibitem[\protect\citeauthoryear{Tomsick et al. }2003]{tomsick03}
         Tomsick J.A. et al., 2003, ApJ, 582, 933        
\bibitem[\protect\citeauthoryear{Wang, Dai \& Lu }2003]{wang03}
         Wang X.Y., Dai Z.G., Lu T., 2003, ApJ, 592, 347  
\bibitem[\protect\citeauthoryear{Wijnands, Homan \& van der Klis }1999]{wijnands99}
         Wijnands R., Homan J., van der Klis M., 1999, ApJ, 526, L33 
\bibitem[\protect\citeauthoryear{Whiteoak \& Green }1996]{whiteoak96}
         Whiteoak J.B.Z., Green A., 1996, A\&AS, 118, 329
\bibitem[\protect\citeauthoryear{Wilson et al. }1998]{wilson98}
         Wilson C.A., Harmon B.A., Paciesas W.S., McCollough M.L., 1998, IAUC 7010  
\bibitem[\protect\citeauthoryear{Wu et al. }2002]{wu02}
         Wu K. et al., 2002, ApJ, 565, 1161     
\bibitem[\protect\citeauthoryear{Wu, Liu \& Li }2007]{wu07}
         Wu Y.X., Liu C.-Z., Li T.-P., 2007, ApJ, 660, 1386      
\bibitem[\protect\citeauthoryear{Xue, Wu \& Cui }2008]{xue08} 
         Xue Y., Wu X.-B., Cui W., 2008, MNRAS, 384, 440         
         
\end{thebibliography}
\end{document}